\documentclass[aps,twocolumn,superscriptaddress,nofootinbib]{revtex4}

\usepackage{amsmath}
\usepackage{amsfonts}
\usepackage{amssymb}
\usepackage{mathdots}
\usepackage{graphicx}
\usepackage{bbm}
\usepackage{color}
\usepackage{pifont}
\usepackage{enumerate}
\usepackage{subfigure}
\usepackage{color}
\usepackage{bbold}
\usepackage [english]{babel}
\usepackage [autostyle, english = american]{csquotes}
\MakeOuterQuote{"}

\begin{document}

\title{Quantum Model Of Spin Noise}


\author{R. Annabestani}
\email[]{rannabes@uwaterloo.ca}
\affiliation{Institute for Quantum Computing, University of Waterloo, Waterloo, Ontario N2L 3G1, Canada}
\affiliation{Department of Physics and Astronomy, University of Waterloo, Waterloo, Ontario N2L 3G1, Canada}
\author{D.G.Cory}
\email[]{dcory@iqc.ca}
\affiliation{Institute for Quantum Computing, University of Waterloo, Waterloo, Ontario N2L 3G1, Canada}
\affiliation{Department of Physics and Astronomy, University of Waterloo, Waterloo, Ontario N2L 3G1, Canada}
\affiliation{Department of Chemistry, University of Waterloo, Waterloo, Ontario N2L 3G1, Canada}
\affiliation{Perimeter Institute for Theoretical Physics, Waterloo, Ontario N2L 2Y5, Canada}
\affiliation{Canadian Institute for Advanced Research, Toronto, Ontario  M5G 1Z8, Canada}
\author{J. Emerson}
\email[]{jemerson@math.uwaterloo.ca}
\affiliation{Institute for Quantum Computing, University of Waterloo, Waterloo, Ontario N2L 3G1, Canada}
\affiliation{Canadian Institute for Advanced Research, Toronto, Ontario  M5G 1Z8, Canada}
\affiliation{Department of Applied Mathematics, University of Waterloo, 200 University Avenue West, N2L 3G1, Waterloo, Ontario, Canada}



\begin{abstract}
Any ensemble of quantum particles exhibits statistical fluctuation{\color{black}s} known as spin noise. {\color{black}Here, we provide a} description of spin noise {\color{black} in the language of} open quantum system{\color{black}s}. The description unifies the signature{\color{black}s} of spin noise under both strong and weak measurements. Further, the model {\color{black}accounts for arbitrary} spin dynamics {\color{black}from an} arbitrary initial state. In all cases we can find {\color{black}both} the spin noise and its time correlation function.
\end{abstract}


\maketitle

\section{Introduction} 
Spin Noise is a signal due to the quantum fluctuations of an ensemble. This {\color{black}phenomenon} has been studied experimentally and theoretically, \cite{Bloch}, \cite{Hahn}, \cite{Ernst}, \cite{Gueron}. Here, we describe an open quantum system approach that provides a simple description of spin noise. This analysis of spin noise {\color{black} may lead} to a clearer understanding of foundational concepts in quantum mechanics such as measurement and fluctuation. The experimental observation of spin noise {\color{black}also} finds application in NMR when the sample has a small number of spins, {\color{black} and/} or a very long relaxation time. \\\\
 Bloch in his original paper in 1946 predicted that even in the absence of any external magnetic field there would still exist a \textit{"resultant moment due to statistically incomplete cancellation"} with a magnitude that scales {\color{black}with the square root of the number of spins} \cite{Bloch}. Sleator $\&$ Hahn \cite{Hahn} observed {\color{black}spin noise} in {\color{black} low temperature} NMR using a high Q superconducting quantum interference device (SQUID resonator). In 1989, Ernst $\&$ McCoy \cite{Ernst} {\color{black}observed spin noise at room temperature in a} high sensitive liquid {\color{black} state} NMR probe. Similarly, Gueron $\&$ Leroy \cite{Gueron} observed spin noise in a sample of water. \\\\
 Spin noise is a signature of any ensemble of quantum systems. There {\color{black}have} been several other observations of spin noise effects including via magnetic resonance force microscopy, spin imaging and optics (\cite{Rugar1},\cite{Rugar2}, \cite{Paoloa}, \cite{Raffi}, and \cite{Alexj}).  {\color{black}Additionally,} Houllt $\&$ Ginsberg and Tropp (\cite{QuantumOrigin},\cite{Tropp}) have given a quantum description of {\color{black} its origin.}\\\\
{\color{black} For spin 1/2 particles} the amplitude of {\color{black} the} spin noise fluctuation grows as {\color{black} the square root of the number of spins}, exists in all directions on the Bloch sphere and has a characteristic correlation time {\color{black}resulting from} the internal Hamiltonian and the relaxation times.\\\\
There are two cases where the spin noise signal is greater than the thermal polarization signal{\color{black}:} a small sample and  a smple with long relaxation time. At equilibrium, the {\color{black}Boltzmann} polarization is $M_{0} \sim N {\color{black}\frac{\hbar \gamma}{2}}\ \tanh(\frac{\hbar \gamma B_{0}}{k T})$ where $\gamma$ is the gyromagnetic ratio of {\color{black}the} spin. The most efficient  detection for a repeated measurement {\color{black} of a free induction decay} is the Ernst angle experiment  with {\color{black} nutation angle $\beta$, set as} $\cos\beta = \exp (-\tau/T_{1})$ {\color{black}where} $\tau$ is the recycle time \cite{ErnstBook}. {\color{black}This} results in {\color{black}a steady state} magnetization {\color{black} of} $ M_{0} \sqrt{(1-\cos\beta)/( 1+\cos\beta)}$ {\color{black} and one can compare it} to the spin noise ($\sim \sqrt{N}{\color{black}\frac{\hbar \gamma}{2}}$) and conclude {\color{black}that} for a small sample, $N< {\color{black}(\sqrt{(1-\cos\beta)/( 1+\cos\beta)}}\epsilon)^{-2}$ {\color{black} and/} or a very long relaxation time, $T_{1}>\tau(\ln[\cos^{-1}[\frac{1- N\epsilon^{2}}{1+ N\epsilon^{2}}]])^{-1}$ {\color{black} where $\epsilon = \tanh(\frac{\hbar \gamma B_{0}}{k T})$}, the spin noise is greater than {\color{black}the} thermal polarization. \\\\   
Here, we apply the theory of open quantum systems to describe {\color{black} the origin and the correlation function of} the spin noise signal. {\color{black} The} analysis shows {\color{black}that} we can model spin noise by {\color{black} separately modeling} the quantum measurement and {\color{black} the} quantum evolution of {\color{black}the} spin {\color{black}system}. {\color{black} First, in section \ref{generalSec}, we outline the general approach and introduce the model. Then,} in section \ref{section1}{\color{black},} we gain physical insight about {\color{black} spin noise} by exploring the case of a totally mixed input state, an  ideal strong measurement and a depolarizing {\color{black}quantum map. This simple yet concrete example allows us to introduce all {\color{black} of} the tools we will need. Following this}, we investigate the case of {\color{black} an} arbitrary quantum evolution acting on a non-interacting ensemble of spins. Finally, we study the effect of {\color{black} weak measurement on the system}.\\
{\color{black}\section{Open Quantum System Model}
\label{generalSec}
In an NMR measurement, an ensemble of spins (sample) is coupled to a bath (environment) and a detection coil. The total Hamiltonian of this system is:
\begin{eqnarray}
\label{TotHam}
\mathcal{H}_{tot}&=& \mathcal{H}_{s} + \mathcal{H}_{B} + \mathcal{H}_{Bs} +  \mathcal{H}_{c} + \mathcal{H}_{sc}\\ \nonumber
\end{eqnarray}
where the first three terms are the spins, the bath and the spin-bath interaction Hamiltonians, and the last two terms are the cavity interaction Hamiltonians. We are interested in the dynamic{\color{black}s} of the spin ensemble {\color{black}alone. S}ince it is interacting with a bath and a measurement apparatus, an open quantum system approach is convenient. In {\color{black}what follows}, we describe an effective  quantum evolution map ( a time snapshot of a propagator) on the N spin ensemble when either just the bath or just the cavity is considered. Then, we combine {\color{black}these} to describe the full evolution. \\
\subsection{N Spins and Bath Interaction}
Consider an initial state with no spins{\color{black}/}bath correlations. Given the time dependent Hamiltonian $\mathcal{H}_{s} + \mathcal{H}_{B} + \mathcal{H}_{Bs}(t)$, this bipartite system evolves under the unitary operator which is the solution of {\color{black}Schr$\ddot{o}$dinger's} equation (\cite{ErnstBook} for {\color{black}a} closed system,
\begin{equation}
\label{unitary}
\rho_{Bs}(t) = U_{Bs}(t). (\rho_{B}(0) \otimes \rho_{s}(0)).U_{Bs}^{\dagger}(t)
\end{equation}
where
$$U_{Bs}(t)= \mathcal{T} e^{ -i \int_{0}^{t} (\mathcal{H}_{s} + \mathcal{H}_{B} + \mathcal{H}_{Bs}(t))\ dt'}.$$
In order to find the reduced evolution operator on the spin ensemble, one can start from Eq.$\ref{unitary}$ and trace over the bath, 
\begin{eqnarray}
\label{QMap}
\rho_{s}(t)&=& Tr_{B}[\rho_{Bs}(t)]\\ \nonumber
&=& Tr_{B}[U_{Bs}(t). (\rho_{B}(0)\otimes \rho_{s}(0)).U_{Bs}^{\dagger}(t)]\\ \nonumber
&=& \Lambda_{t}[ \rho_{s}(0)].
\end{eqnarray}
The quantum evolution map, $\Lambda_{t}$, is not generally a unitary evolution. This is the distinction between a closed and an open system. Generally, if the bath interaction is Markovian, then the dynamic{\color{black}s} of the open quantum system follow a master equation( \cite{ErnstBook} $\&$ \cite{OQSBook}) 
\begin{equation}
\label{master}
\frac{\partial \rho_{s}(t)}{\partial t}= -i [\mathcal{H}_{s}, \rho_{s}(t)]+ \hat{\Gamma}\ [\rho_{s}(t)]
\end{equation}
where the evolution depends on both the coherent evolution, $-i[\mathcal{H}_{s},\cdot]$, and a dissipater, $\Gamma[\cdot]$, which describe the effective result of coupling to the bath. This term leads to decoherence or relaxation and drives the system towards its equilibrium state. The defining characteristic of the quantum map, $\Lambda_{t}$, is that it takes a density matrix to a density matrix for an initially uncorrelated state of the spin and the bath. Such a map is {\color{black}called} completely positive  {\color{black}and} trace preserving (CPTP), Fig.\ref{Evolution}. \\ 
\begin{figure}[h]
\center
\includegraphics[width=0.45 \textwidth]{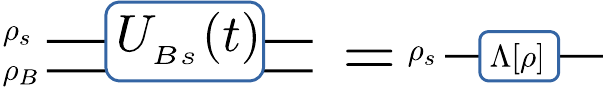}
\caption{\label{Evolution} The reduced time evolution operator (CPTP map) on an ensemble of spins coupled to a bath.}
\end{figure}\\ 
For completeness we briefly describe how to connect these two descriptions. In a master equation, the relaxation operator $\hat{\Gamma}$ is induced by the bath coupling. The interaction Hamiltonian can be written as:
$$\mathcal{H}_{Bs}= \sum\limits_{i} B_{i}(t)\otimes A_{i}$$
where the operators $A_{i}$ are acting on the spin system and the operators $B_{i}(t)$ are fluctuating randomly and are acting on the bath system. One can find the bath time correlation function
\begin{equation}
R^{ij}(\tau)= \overline{B_{i}^{\dagger}(t)B_{j}(t-\tau)}
\end{equation}
from which the spectral density of noise is known, $J^{ij}(\omega)= \int d\tau e^{i\omega t} R^{ij}(\tau)$. Then, under some assumptions (\cite{ErnstBook} $\&$ \cite{OQSBook}), one can find the relaxation superoperator,
\begin{equation}\nonumber
\hat{\Gamma}\ [\rho_{s}(t)] = \sum\limits_{i,j,\omega} J_{ij}(\omega)[A_{j}(\omega)\rho_{s}(t)A_{i}^{\dagger}(\omega)-A_{i}^{\dagger}(\omega)A_{j} (\omega)\rho_{s}(t)+h.c]
\end{equation}
where $A(\omega)$ is the component of $A$ in the frequency domain. Given this last relation for $\hat{\Gamma}[\rho_{s}]$, the solution of the master equation in Eq.(\ref{master}), is the same as the quantum evolution map defined in Eq.(\ref{QMap}) under Markovian interaction.
\subsection{ Cavity Interaction }
We can also find an effective map for the coupling of the spins to the cavity{\color{black}/}coil, $\mathcal{H}_{c}$ and $\mathcal{H}_{sc}$. This system evolves unitarily 
\begin{equation}\nonumber
\rho_{sc}(t)= U_{sc}. (\rho_{s} \otimes |\psi\rangle _{c}\langle \psi|). U_{sc}^{\dagger}  
\end{equation} 
where $U_{sc}= \exp(- i (\mathcal{H}_{c}+ \mathcal{H}_{sc})\ t)$. 
For our analysis, the cavity does not distinguish between spins and the spins only couple to a single {\color{black}mode}. This is described by the Tavis-Cumming Hamiltonian, (\cite{Tavis})
\begin{eqnarray}
\mathcal{H}_{c}+ \mathcal{H}_{sc} = \omega_{c} \ a a^{\dagger}+ g \ J_{x} (a+a^{\dagger}) .\nonumber
\end{eqnarray}
Here $J_{x}= \sum\limits^{N}_{i=1} S^{(i)}_{x}$ is the x component of the total spin angular momentum, and $a$ and $a^{\dagger}$ are the ladder operators. According to this model, the detection coil does not distinguish spins and in a measurement, the net magnetization of the whole ensemble is recorded{\color{black}, given by} $m= \sum\limits_{i=1}^{N}s^{(i)}$ with $s \in \{+ \frac{1}{2},- \frac{1}{2}\}$. So, a measurement leaves the spin ensemble in a totally symmetric submanifold with net magnetization $m$. {\color{black} To see the effect of measurement explicitly, we note that the cavity is coupled to additional degrees of freedom that produce the observed measured outcomes (e.g, electronics). So effectively, the spin system couples to the measurement device ($c'$) via the cavity interactions and once the measurement is completed the cavity is left in its initial state. This allows us to drop the cavity from the model. If the detection has} an accuracy of one single spin flip, the possible measured outcomes are $m\in  \{ -\frac{N}{2},-\frac{N}{2}+1, ..., \frac{N}{2}\}$ and correspondingly {\color{black} the measurement device Hilbert space is spanned by} an orthonormal basis $\{|m\rangle\}$. {\color{black} For the evolved state $\rho_{sc'}(t)$, tracing over the measurement device gives}
\begin{eqnarray}
\label{Sys-Cavity}
Tr_{c'}[\rho_{sc'}(t)] &=& Tr_{c'}[U_{sc'}. (\rho_{s}(0) \otimes |\psi\rangle \langle \psi|{\color{black})}. U_{sc'}^{\dagger}]\\ \nonumber
\rho_{s}(t)&=& \sum\limits_{m} \langle m|U_{sc'}|\psi\rangle. \rho_{s}(0). \langle m|U_{sc'}|\psi\rangle^{\dagger} \\ \nonumber
&=& \sum\limits_{m} \mathcal{M}_{m}. \rho_{s}(0). \mathcal{M}_{m}^{\dagger} 
\end{eqnarray} 
where $\mathcal{M}_{m} = \langle m| U_{sc'}|\psi\rangle$ is defined as the measurement operator assigned to the measurement outcome $m$. {\color{black}Here, the partial trace of any operator $\hat{O}$ is defined as $Tr_{B}[\hat{O}_{AB}]= \sum\limits_{b} (\mathbb{1}\otimes \langle b|)\hat{O}_{AB}( \mathbb{1}\otimes | b\rangle)$. According to Eq.(\ref{Sys-Cavity})}, the effect of {\color{black} the} interaction with the detection coil appears as an effective quantum map $\mathcal{E}[\rho] = \sum\limits_{m} \mathcal{M}_{m}. \rho_{s}. \mathcal{M}_{m}^{\dagger}$ on the spin ensemble. It is easy to check that $\mathcal{E}[\rho]$ is also a CPTP map and hence, $\sum\limits_{m} \mathcal{M}^{\dagger}_{m} \mathcal{M}_{m} = \mathbb{1}$.\\\\ 
Notice {\color{black}that} we considered an initial pure state $|\psi\rangle$ for the {\color{black}measurement device}. One can generalize this argument for any initial mixed state, $\rho_{c}(0)$,  because it can be written as a {\color{black}convex} combination of pure states and all {\color{black} of} the maps in the presented model are linear.\\\\
In a quantum measurement there is a trade off between the amount of information obtained and the amount of disturbance introduced {\color{black}in} the system. Say the detection coil measures the classical value $m_{0}$, then {\color{black}the} spin ensemble's state conditioned on the knowledge $m_{0}$ is updated to \cite{Nielson}
$$ \rho^{|m_{0}} =\frac{\mathcal{M}_{m_{0}}. \rho_{s}. \mathcal{M}_{m_{0}}^{\dagger}}{P(m_{0})}$$
where $P(m_{0})=Tr[\mathcal{M}^{\dagger}_{m_{0}} \mathcal{M}_{m_{0}} \rho_{s}]$ is the probability that such an event occurs.
\begin{figure}[h]
\center
\includegraphics[width=0.5 \textwidth]{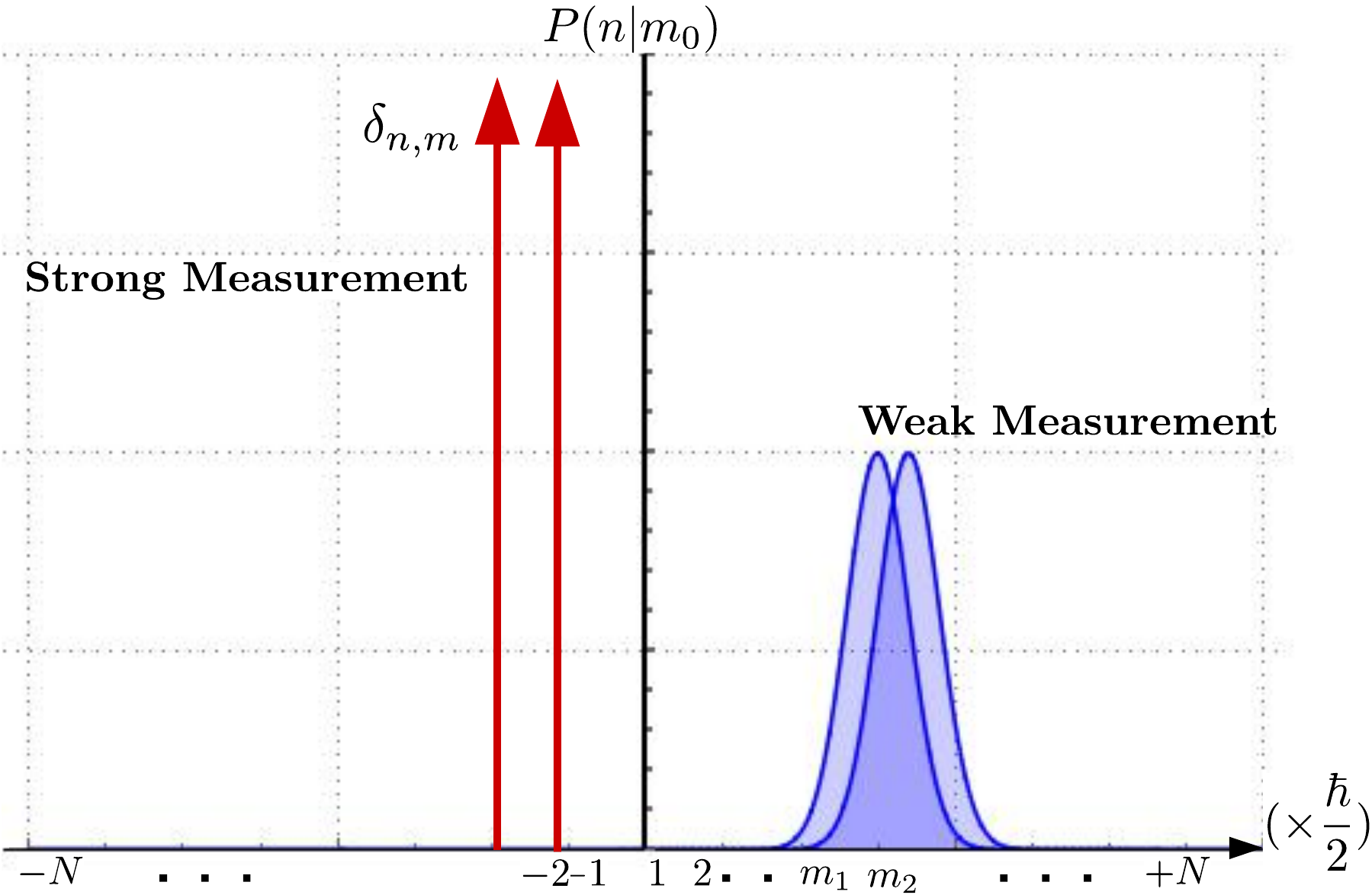}
\caption{\label{POVM-Figure} The conditional probability distribution of a strong measurement and a weak measurement {\color{black} are compared}. In a strong measurement, the updated density matrix collapses sharply to the submanifold $m_{0}$ {\color{black}and this leads to a} delta function distribution (red). Whereas, in a weak measurement, the density matrix is less disturbed and it collapses to an area centered at $m_{0}$. {\color{black}So, the corresponding probability distribution has a} finite width (blue). The horizontal axis is in $\frac{\hbar}{2}$ unit {\color{black}and} represents the net magnetization of the ensemble.}
\end{figure}
 In the case of a strong measurement, the ensemble magnetization, $m_{0}$, is known with certainty. Therefore, the spin ensemble density matrix collapses (disturbance) to the $m=m_{0}$ manifold only, and if we make a second measurement immediately {\color{black}afterwards}, the outcome $m_{0}$ is reproduced. In other words, the conditional probability distribution of the second measurement is a delta function, i.e, $P(n|m_{0}) = \delta_{n, m_{0}}$. In the case of a weak measurement, the measurement apparatus is less precise and the spin ensemble state collapses not only to the $m=m_{0}$ manifold but also to the other neighboring manifolds, $n\neq m_{0}$. So, if we {\color{black}immediately} make another measurement, the outcome $m_{0}$ may not be reproduced. In other words, the conditional probability distribution  $P(n|m_{0})$ could be a distribution function with mean value $m_{0}$ and a width $w$ which is in inverse relation with the accuracy of the measurement device (Fig.\ref{POVM-Figure}). We will provide a more detailed model of a strong and a weak measurement in sections \ref{section1} and \ref{section2}. 
\subsection{N Spins Coupled to the Bath and the Cavity}
So far, we have considered the effect of coupling to the measurement apparatus and the reduced quantum evolution map on the spin ensemble as two independent processes. However, in an NMR measurement these two processes occur simultaneously (Eq.\ref{TotHam}). So,
$$ U_{Bsc}(T)= \mathcal{T} e^{ -i \int_{0}^{T} \mathcal{H}_{tot}(t') \ dt'}.$$
The various contributions of $\mathcal{H}_{tot}$ do not in general commute at all times and so the formal solution is not practically helpful. One can discretize the total evolution time, $T= n\ t$, in which $t$ is small enough to allow a first order approximation. Then, for a short time evolution $t$, the first order of the Magnus expansion (\cite{Waugh}) is
\begin{eqnarray}
U_{Bsc}(t)&\approx &  e^{ -i \int_{0}^{t}(\mathcal{H}_{s}+ \mathcal{H}_{B}+\mathcal{H}_{Bs}(t')) dt} e^{-i (\mathcal{H}_{c}+ \mathcal{H}_{sc})t}\\ \nonumber
 &= &  U_{sc}(t)U_{Bs}(t)
 \end{eqnarray}
 \begin{figure}[h]
\center
\includegraphics[width=0.5 \textwidth]{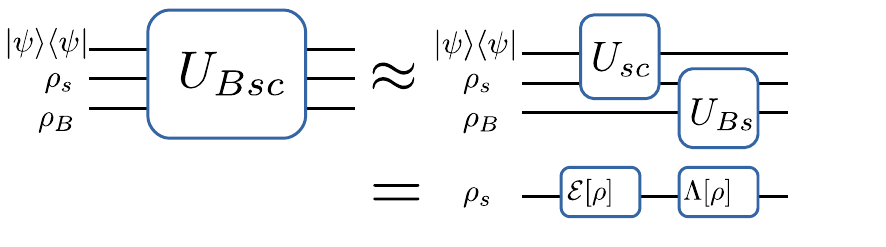}
\caption{\label{3Party} A first order approximation of time evolution of a bath-spin ensemble-cavity system is presented.}
\end{figure}\\
For example, in the case of $\mathcal{H}_{s} = \omega_{0} J_{z}$ and the Tavis-cumming model for interaction with the cavity, this approximation is valid if $t\ll \frac{1}{\sqrt{\omega_{0} g}}$. In this first order approximation, the effective quantum evolution map for the short time period $t$ on the spin ensemble is: 
 \begin{eqnarray}
  \rho_{s}(t)&=& \hat{\mathcal{S}}_{t}[\rho_{s}]\\ \nonumber
  &=& Tr_{Bc}[ U_{Bsc}. \rho_{Bsc}(0). U_{Bsc}^{\dagger}]\\ \nonumber
  &\approx & Tr_{Bc}[ (U_{Bs}.U_{sc}). \rho_{Bsc}(0). (U_{Bs}.U_{sc})^{\dagger}]\\ \nonumber
  &=& \Lambda[ \mathcal{E}[\rho_{s}(0)]]
 \end{eqnarray}
where $\rho_{Bsc}(0)=\rho_{B}(0) \otimes \rho_{s}(0) \otimes |\psi\rangle \langle \psi|$.
Therefore, a quantum evolution map on the spin ensemble, $\hat{\mathcal{S}_{T}}$, can be approximated by a sequence of measurement-evolution processes as schematically is shown in Fig.\ref{sequence}.
\begin{figure}[h]
\center
\includegraphics[width=0.5 \textwidth]{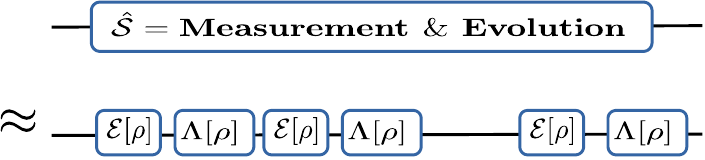}
\caption{\label{sequence} The effective time evolution operator $\hat{\mathcal{S}}$ on a spin ensemble is approximated by a sequence of measurement-evolution processes.}
\end{figure} \\\\
In the following sections, we apply this model to the examples of both strong and weak measurements under the evolution of {\color{black}a} collective depolarizing map or any arbitrary CPTP map on individual spins. In each cases, we find the spin noise and its correlation function.}
\section{Strong Measurement Model}
\label{section1}
  Suppose we have $N$ identical spin half particles and we have no information about their spin orientation. So, at $t=0$, the density matrix $\rho_{0}=\frac{\mathbb{1}}{2^{N}}$ describes "our knowledge" about the system which is maximal ignorance. Now, {\color{black} according to the model presented in previous section}, we make a series of strong measurements on the system by which we obtain information about the collective magnetization, $\textbf{M}$. Between two subsequent measurements, there is a time interval $\delta t$ during which the system evolves under a quantum {\color{black}evolution map} $\Lambda_{\delta t}$.
  Without loss of generality, we assume the collective measurements are along the z axis. Of course, NMR detection is in the $ x-y$ plane, but, for this analysis the direction is of no importance. At $t=t_{n}$, the recorded data, $\textbf{M}(t_{n})${\color{black},} are eigenvalues of {\color{black}the z component of} the total {\color{black}spin} angular momentum, $\textbf{J}_{z}= \sum\limits^{N}_{i=1}\ S^{(i)}_{z}$. This choice of collective measurement is not {\color{black}the} common {\color{black} one in NMR}, {\color{black}usually} $N\langle S_{z} \rangle$ is used as the ensemble signal. However, in order to see the spin noise effects, one needs to keep track of what has been learned about the ensemble in each measurement {\color{black}rather than just the mean value}. Therefore we do the analysis in the total angular momentum space. This has been used before \cite{Poulin}.\\ \\
 The action of a strong measurement is {\color{black}described} by a set of projection valued measure (PVM) operators, {\color{black}which we denoted as} ${\color{black}\{\mathcal{M}_{m}=}\Pi_{m}\}$ {\color{black}and} are given by
\begin{eqnarray}
\textbf{J}_{z}&=&\sum\limits_{m} m \ \Pi_{m} \\ \nonumber
\Pi_{m}&\equiv& \sum_{j=|m|}^{N/2} \sum_{a=1}^{A_{j}}\  |j, m, a \rangle \langle j, m, a | \nonumber.
\end{eqnarray}
where $\hbar=1$. {\color{black}Here, $|j, m,a\rangle$ are degenerate eigenstates of the total spin angular momentum $\vec{J}= \sum\limits_{i=1}^{N} \vec{S}^{(i)}$ as well as its z component $\textbf{J}_{z}$ operator. For $N$ spin half particles, $j= j_{0}, j_{0}+1, ...,N/2$ where $j_{0}= 0\ ( 1/2)$ if N is even (odd). For each total spin angular momentum's eigenvalue, $j$, the collective magnetization in {\color{black}the} z direction is $m= -j,j+1,..., j$, and, the state degeneracy label is $a=1,2,..., A_{j}$ where $A_{j}=\left(\begin{array}{c}N\\ \frac{N}{2}+ j\end{array}\right)-\left(\begin{array}{c}N\\ \frac{N}{2}+ j+1\end{array}\right)$[18]. These eigenstates span the whole Hilbert space and form a basis for an ensemble of spins. It is common to consider $j$ as the principle quantum number and, $m$ as the second quantum number. However, mathematically it is equivalent to consider $m$ as the principle number, $m \in \{ -N/2, -N/2+1, ..., N/2 \}$ and $|m|\leq j \leq N/2$ as the second quantum number which is the case in our notation.  Note, by this definition,  $\lbrace\Pi_{m}\rbrace$ satisfies the conditions of Projective Value Measure (PVM) operators, i.e, 
 $ \Pi_{m}.\Pi_{n}= \delta_{mn} \Pi_{m}$ and $\sum\limits_{m} \Pi_{m} = \mathbb{1}$. An example of a strong measurement on a single spin is Stern-Gerlach experiment where the measurement operators are $\Pi_{+}= |\uparrow\rangle \langle \uparrow|$ and $\Pi_{-}= |\downarrow\rangle \langle \downarrow|$ which are orthogonal projective operators corresponding to the outcome "up" and "down". Here, $\Pi_{m}$ are the generalized form for {\color{black}an} $N$ spin projective measurement when the detection coil has the precision of one single spin.}\\ \\
 The first measurement at $t_{1}$, results in outcome $m_{1}\in[-\frac{N}{2},\frac{N}{2}]$ which occurs with probability $P(m_{1}; t_{1})$. This probability is a binomial (semi-Gaussian) distribution with zero mean and $\sqrt{N}$ standard deviation, because 
\begin{eqnarray}
\label{statistics}
P(m_{1};t_{1})&=&Tr[\Pi_{m_{1}}.\rho_{0}]= \frac{Tr[\Pi_{m_{1}}]}{2^{N}} \\ \nonumber
E[\textbf{M};t_{1}]&=&Tr[J_{z}.\rho_{0}]=0\\ \nonumber
\sigma[\textbf{M};t_{1}]&=&\sqrt{ Tr[J_{z}^{2}.\rho_{0}]- (Tr[J_{z}.\rho_{0}])^2}= \frac{1}{2}\sqrt{N}\\ \nonumber
\end{eqnarray} 
This result matches what we intuitively expect. {\color{black} Each spin has magnetization} $s_{i} \in \lbrace +\frac{1}{2}, -\frac{1}{2} \rbrace$, and, in each measurement shot, we take $N$ samples from a distribution $P(s)$ with {\color{black}a} width of $\frac{1}{2}$.{ \color{black} Therefore, according to the central limit theorem, the collective magnetization} $m=\sum^{N}\limits_{i=1} s_{i}$ itself is a random variable whose distribution is Gaussian with width of $\frac{\sqrt{N}}{2}$. Because the spins are indistinguishable, $Tr[\Pi_{m}]$ counts the number of configurations that all result in $m$ net magnetization and therefore $P(\textbf{M};t_{1})$ is a binomial distribution. \\\\
Once we learn the system, we must update its density matrix according to "our knowledge" of the outcome. So, given the outcome $m_{1}$, the state update rule \cite{Nielson} {\color{black}dictates that}
 \begin{equation}
 \label{update}
 \rho^{|\textbf{M}=m_{1}}= \frac{\Pi_{m_{1}}.\rho_{0}.\Pi_{m_{1}}}{P(m_{1};t_{1})}= \frac{\Pi_{m_{1}}}{Tr[\Pi_{m_{1}}]}.
 \end{equation}
The state (\ref{update}) evolves under a quantum {\color{black}map} $\Lambda$ during the time interval $\delta t$ after which the next measurement takes place. {\color{black} As an example}, we consider a collective depolarizing {\color{black}map} where with probability $(1-\lambda)= \text{exp}[-\delta t/ T]$ the quantum state is preserved and with probability $\lambda$ it turns to a fully mixed state. The characteristic time $T$ is a function of the depolarizing strength. Physically, a depolarizing {\color{black}map} could be a result of a relaxation process in the system and mathematically is given by   
  \begin{equation}
\Lambda[\rho]= (1-  \lambda) \  \rho + \lambda \ \frac{\mathbb{1}}{2^{N}} .
\end{equation}
 Now, in the second step, the evolved state $\Lambda[\rho^{|m_{1}}]$ is measured and outcome $m_{2}$ is obtained whose probability is given by $P(m_{2};t_{2}|m_{1})= (1-\lambda)\ \delta_{m_{1}, m_{2}} + \lambda\ P(m_{2};t_{1})$. This  $P(\textbf{M};t_{2}|m_{1})$ will be again a semi-Gaussian distribution with a conditional mean and conditional standard deviation 
 \begin{eqnarray} 
 \label{Correlation}
 E(\textbf{M};t_{2}|m_{1})&=&  (1-\lambda)\  m_{1} \\ \nonumber
\sigma(\textbf{M};t_{2}|m_{1})&=& \sqrt{\lambda\ (\frac{N}{4}+ (1-\lambda)\  m^{2}_{1})}\\ \nonumber
\end{eqnarray}
Thus, the second measurement statistics {\color{black}are} correlated with the first measurement outcome $m_{1}$. This correlation does not last forever and is limited by the relaxation time of the dissipative system, $T$. For instance, if we record data so slowly, $\delta t >> T$( or $\lambda\rightarrow 1$), each measurement data $m_{k}$ is sampled from a fixed distribution $P(\textbf{M};t_{1})$ with zero mean and $\frac{1}{2}\sqrt{N}$ standard deviation and there will be no correlation between data, Eq.(\ref{Correlation}). In another extreme case, when we record data quickly, $\delta t << T$, then $1-\lambda\approx 1- \frac{\delta t}{T}$ and the system does not evolve, hence, the data is repeatable, which is a property of a projective measurement. In non-extreme regimes, when $\delta t < T$, the data is sampled from semi- binomial distributions whose mean and variance are fluctuating from one measurement to another.\\\\
 After a long data acquisition a list of {\color{black} outcomes} $\{ m_{1}, m_{2}, ...., m_{k}\}$ {\color{black} is obtained} which constructs {\color{black}the} spin noise signal. The spin noise is the {\color{black} net} magnetization {\color{black} of an ensemble} whose fluctuating value is bounded by  $\frac{N}{2}$ and -$\frac{N}{2}$. At step $k$th, $m_{k}$ is a random variable sampled from semi-Gaussian distribution  $P(\textbf{M};t_{k}|m_{k-1},...,m_{2}, m_{1})$ whose mean and variance {\color{black}are} correlated with previous recorded data. For the particular choice of a depolarizing {\color{black}map},  using inductive reasoning, we obtain {\color{black}that} the the joint probability distribution between any two data points {\color{black}is}
\begin{eqnarray}
\label{JointDist}
 P(m_{i};t_{i}, m_{j}; t_{j})&=& (1-\lambda)^{i-j}\ \delta_{m_{i}, m_{j}}\ P(m_{j};t_{j}) \\ \nonumber &+& \eta_{i-j} \ P(m_{i};t_{i})\ P(m_{j};t_{j})\nonumber
\end{eqnarray}
where $\eta_{k}= \lambda+ (1-\lambda)\ \eta_{k-1}$ and $\eta_{0}=0$ and $P(m; t_{i})= P(m;t_{1})= Tr[\Pi_{m}.\rho_{0}]$. Relation.(\ref{JointDist}) indicates that, with the probability of $(1-\lambda)^{k} \sim e^{-t_{k}/T}$,  the two measurements separated by $t_{k}= k\ \delta t$, are perfectly correlated and with the probability of $\eta_{k}$, they are two independent random variables. In other words, the closer the two measurements are {\color{black}in time}, the more likely that their distributions {\color{black}are} correlated. Given Eq.(\ref{JointDist}), one can compute the covariance function as a measure of {\color{black}the} correlation,
\begin{eqnarray}\nonumber
\label{PVMCorr}
R(k)&\equiv & E(\textbf{M};t_{k+i}, \textbf{M},t_{i}) -E(\textbf{M};t_{k+i})E( \textbf{M},t_{i})\\
&=&\frac{N}{4}\ e^{-t_{k}/T}
\end{eqnarray}
where the expectation values are calculated using $E(X;t_{1})= \sum_{x}x\ P(x;t_{1})$ and $E(X;t_{i},Y;t_{j})= \sum_{x,y} x\ y\ P(x;t_{i}, y, t_{j})$ and we assumed {\color{black}an initially fully mixed state.}\\\\ 
This analysis {\color{black}has considered} a collective evolution $\Lambda$  and a collective measurement $\Pi_{m}$ over an ensemble {\color{black}where the} collective measurement preserves coherences within the subspace $m$.   \\
\subsection{Arbitrary Quantum Map $\Lambda$ for non-interacting spins}
\label{subA} 
We can further generalize the description by extending it to any arbitrary CPTP quantum map acting on individual spins. More precisely, {\color{black} suppose the} spins are not interacting with each other{\color{black}, that} there is no field inhomogeneity and also no variation of {\color{black}the} $B_{1}$ field, {\color{black}and, that} each individual spin interacts with its own bath. {\color{black}Therefore}, spins are indistinguishable to the environment and one can model the ensemble quantum evolution as $\Lambda = \Phi ^{\otimes N}$ where $\Phi$ is a CPTP map on a single spin. {\color{black}In this picture, each spin is an open quantum system.}\\\\
{\color{black}As before, consider a totally mixed initial state for each spin}, $\rho_{0}= (\frac{\mathbb{1}}{2})^{\otimes N}$, and make a strong measurement along the z axis. {\color{black} Upon the measurement with outcome $m$, there are $\frac{N}{2}+m$ number of spins with up orientation and $\frac{N}{2}-m$ with down orientation. So, the} measurement statistics {\color{black}are} given by
\begin{eqnarray}
P(m;t_{1})&=&\left(\begin{array}{c}N\\ \frac{N}{2}+ m\end{array}\right)(\frac{1}{2})^{\frac{N}{2}+ m} (\frac{1}{2})^{\frac{N}{2}- m}.\\ \nonumber
\end{eqnarray}
{\color{black}
 The PVM operator given in Eq.(1) can also be expanded in the tensor product basis as:}  $$\Pi_{m}= \sum\limits_{p}\hat{ \mathcal{P}}_{p} [ |\uparrow \rangle\langle \uparrow|^{\otimes \frac{N}{2}+ m} \otimes |\downarrow\rangle\langle \downarrow |^{\otimes \frac{N}{2}- m}].$$ 
{\color{black} Here, since the spins are indistinguishable, there is a sum over all possible spin permutations that results in net spin magnetization $m$.} So, $p\in\lbrace 1, ...,\left(\begin{array}{c}N\\ \frac{N}{2}+ m\end{array}\right)\rbrace$. Upon {\color{black}recording the classical value} $m_{1}$, the density matrix is updated to
\begin{eqnarray}
\label{updateState}
\rho^{|m_{1}}&=&\frac{ \Pi_{m_{1}}. \rho_{0}. \Pi_{m_{1}}}{P(m_{1};t_{1})}\\\nonumber
&=& \frac{\sum\limits_{s} \hat{ \mathcal{P}}_{s} [ |\uparrow \rangle\langle \uparrow|^{\otimes \frac{N}{2}+ m_{1}} \otimes |\downarrow\rangle\langle \downarrow |^{\otimes \frac{N}{2}- m_{1}}]}{2^{N}P(m_{1};t_{1})}.
\end{eqnarray}
This updated state evolves under $\Lambda$ which means that each spin evolves under $\Phi$. An example of a single qubit CPTP map $\Phi$ would be a rotation around axis $\hat{r}_{1}$, a relaxation around axis $\hat{r}_{2}$ and a dephasing around  axis $\hat{r}_{3}$ on the Bloch sphere. {\color{black} In general, the action of a map $\Phi$ on the spin basis can be written as}
$$
\Phi[|\uparrow \rangle\langle \uparrow|]= (1-\alpha)\ |\uparrow \rangle\langle \uparrow| + \alpha \ |\downarrow \rangle\langle\downarrow|+ \text{Off diagonal}$$
$$\Phi[|\downarrow\rangle\langle \downarrow |]= (1-\beta)\ |\downarrow \rangle\langle \downarrow|+ \beta\ |\uparrow \rangle\langle \uparrow|+\text{Off diagonal} 
$$
where $\alpha$ and $\beta$ are variables which are determined by the {\color{black}map's} parameters such as evolution time $\delta t$, frequency $\omega$, relaxation and dephasing rates and the directions $\hat{r}_{1},\hat{r}_{2},\hat{r}_{3}$.\\\\
The second measurement on $\Lambda[\rho^{|m_{1}}]$ results in outcome $m_{2}$ which occurs with probability 
\begin{eqnarray}
\label{SecondProb}
P(m_{2};t_{2}|m_{1})=\sum^{\frac{N}{2}+m_{1}}\limits_{k=0}\sum^{\frac{N}{2}-m_{1}}\limits_{l=0}&&\lbrace\text{Bin}(\frac{N}{2}+m_{1}, k, \alpha)\\\nonumber
&& \text{Bin}(\frac{N}{2}-m_{1}, l, \beta)  \delta_{_{m_{2}- m_{1}, l-k}}\rbrace \\ \nonumber
\end{eqnarray}
where $\text{Bin}( n, k, p)= \left(\begin{array}{c}n\\ k\end{array}\right) p^{k} (1-p)^{n-k}$. This new distribution is again a binomial with {\color{black} mean value}
$$ E(\bold{M}, t_{2}| m_{1})= m_{1}(1-(\alpha+\beta)) +(\frac{N}{2}) (\alpha - \beta) $$
and {\color{black} standard deviation}
\begin{eqnarray}\nonumber
\sigma(\textbf{M};t_{2}|m_{1})&=& \sqrt{\frac{N}{2}(\alpha(1-\alpha)+ \beta (1-\beta)}\\\nonumber
&+&\sqrt{m_{1}(\alpha(1-\alpha)- \beta (1-\beta)}\nonumber. 
\end{eqnarray} \\
 As {\color{black}the above} relations indicate, depending on the evolution map's parameters, $\alpha$ and $\beta$, the statistics of the noise {\color{black}are} different. Nevertheless, the spin noise magnitude still scales with $\sqrt{N}$ and exhibits a time correlation. Notice, it is not necessary to consider an open system interacting with an environment to see the spin fluctuation. For example, even in the case of simple unitary evolution where $\alpha= \beta = \sin^{2}[\omega \ \delta t]$, {\color{black}these} correlated fluctuation{\color{black}s} exist.\\\\
  {\color{black}In order to find the correlation function, we need to know the joint probability distribution, Eq.(\ref{PVMCorr}).} In the particular choice of a totally mixed input state, after each measurement, the updated density matrix is $\Pi_{m_{k}}/Tr[\Pi_{m_{k}}]$. {\color{black} Therefore}, $P(m_{k};t_{k}|m_{{\color{black}k}-1})= P(m_{2};t_{2}|m_{1})$ for all $t_{k}$, and hence, the joint probability distribution of any two data points is:\\
\begin{eqnarray}
\label{JointDistPhi}
P(m_{i};t_{i},\  m_{j}; t_{j})&=&\sum\limits_{l_{i-j},...,l_{i-1}} P(m_{i};t_{i}|l_{i-1})\\ \nonumber
&& \times...\times \ P(l_{i-j};t_{i-j}|m_{j}) P(m_{j};t_{j})\nonumber.
\end{eqnarray}\\ 
{\color{black} Substituting Eq.(\ref{SecondProb}) into} relation (\ref{JointDistPhi}) gives us an analytic expression for the joint probability distribution and in the large ensemble limit and for the totally mixed input state, one can approximate each $P(m_{k}, t_{k}|m_{k-1})$ with a Gaussian distribution whose mean and variance are fluctuating from one measurement to the next. \\\\
\newpage
 \subsubsection{Arbitrary Initial State}
 \label{subB}
{\color{black} In this section, we consider} $N$ identical and non interacting spins, $\rho_{0}= \varrho^{\otimes N}$ where $\varrho$ is an arbitrary single spin density matrix {\color{black} that is} expanded as: 
\begin{equation}
\label{rho-general}
\varrho = a \ |\uparrow \rangle\langle \uparrow| + (1-a) \ |\downarrow \rangle\langle\downarrow|+ b \ |\uparrow \rangle\langle\downarrow| + b^* \ |\downarrow \rangle\langle\uparrow|.
\end{equation}
The first measurement on this ensemble results in the statistical distribution
\begin{eqnarray}
\label{prob}
P(m_{1};t_{1})&=&\left(\begin{array}{c}N\\ \frac{N}{2}+ m\end{array}\right)(a)^{\frac{N}{2}+ m_{1}} (1-a)^{\frac{N}{2}- m_{1}}.\\ \nonumber
\end{eqnarray}
This distribution does not distinguish $\varrho$ from a diagonal state $\tilde{\varrho}= a \ |\uparrow \rangle\langle \uparrow| + (1-a) \ |\downarrow \rangle\langle\downarrow|$ since the measurement is along the z axis. Therefore, it is sufficient to consider $\tilde{\varrho}$ as an arbitrary initial state. Upon the strong measurement given by $\Pi_{m}$, the state update rule {\color{black}implies that} \\
\begin{eqnarray}
\rho^{|m_{1}}&=& \sum\limits_{s}(a)^{\frac{N}{2}+ m_{1}} (1-a)^{\frac{N}{2}- m_{1}}\\ \nonumber
&\times&\frac{  \hat{ \mathcal{P}}_{s} [ |\uparrow \rangle\langle \uparrow|^{\otimes \frac{N}{2}+ m_{1}} \otimes |\downarrow\rangle\langle \downarrow |^{\otimes \frac{N}{2}- m_{1}}]}{P(m_{1};t_{1})}.
\end{eqnarray} 
By replacing $P(m_{1};t_{1})$ {\color{black} with Eq.\ref{prob},} we see that the above state is identical to the updated state (\ref{updateState}) where the experiment started from a mixed state. Despite the fact that the first measurement statistics differentiate an arbitrary initial state ($\varrho$ or $\tilde{\varrho}$) from an identity state ($\mathbb{1}/2^N$), their corresponding updated states are no longer distinguishable to the subsequent measurement-evolution processes. As a result, except for the first data point, the statistical fluctuation{\color{black}s} of spin noise {\color{black}are} the same whether we start from a mixed state or from an arbitrary initial state.
\section{Weak Measurement Model}
    \label{section2} 
 {\color{black} In an NMR measurement, it is too idealistic to assume that the detection process can resolve a single spin. If we relax this assumption, the projective measurement operators, $\mathcal{M}_{m}= \Pi_{m}$, no longer describe the action of {\color{black}a} measurement. One needs to assign a width of precision to the measurement apparatus which results in an overlap between the different subspaces (Fig.\ref{POVM-Figure}). Therefore, once the data $m_{0}$ is recorded, the spin ensemble density matrix collapses not only to the $m_{0}$ subspace but also to other subspaces with $l\neq m_{0}$.} The most general type of measurement are mathematically modeled by positive valued operator measure (POVM) operators, {\color{black} $E_{m}$, which {\color{black}result in} measurement statistics $P(m)=Tr[E_{m}\rho]$, \cite{Nielson}. A PVM is a special case of this.} Therefore, we adapt the spin noise model by relaxing the assumption of a strong measurement to a weak measurement and defining POVM elements, {\color{black}$E_{m}= \mathcal{M}_{m}^{\dagger}\mathcal{M}_{m}$, as a sum of PVM operators}, 
\begin{equation}
E_{m}= \sum\limits^{N/2}_{l=-N/2} D(m,l)\ \Pi_{l}
\end{equation}
where $D(m,l)$ is a two variable function {\color{black}whose} form is limited by physical constraints{\color{black}:} \\\\
\begin{enumerate}
\item The measurement is trace preserving. So,
$$
\sum\limits_{m} E_{m}= \mathbb{1} \ \Rightarrow\  \mbox{ for each}\  l \hskip 0.25 cm \sum_{m} D(m, l)=1.
$$
This means {\color{black}that} $D$ is certainly a distribution relative to $m$, but it does not have to be a distribution relative to $l$. This condition {\color{black} becomes particularly important} when we get close to the boundaries $\pm \frac{N}{2}$.\\\\
\item Since the detector records data $m$ as the outcome, we expect $D$ to have its maximum value at $l=m$. So, $$\max\limits_{l} D(m,l)=D(m,m).$$\\\\
\item In a weak measurement, the measurement outcome is less reliable; if the measurement apparatus records $m$, {\color{black} there is a} probability $D(m,l)$ that the updated system collapses to other subspaces with $l\neq m$. One expects {\color{black} the further apart $l$ and $m$ are, the less likely it is to collapse into the $l$} subspace. {\color{black}Thus}, $D(m,l)$ should decrease as $|l-m|$ increases and its width should be {\color{black}inversely propotioned to} the reliability of the measurement device, $1/w$.\\\\
\item $D$ {\color{black} need not be} a symmetric function. For instance, we know it must be a distribution relative to $m$ but {\color{black} it need not} have restriction relative to $l$. So, in general $D(a,b) \neq D(b,a)$.\\\\
\end{enumerate}  
Considering the above constraints, we model the function $D(m,l)$ by a semi-Gaussian distribution:
\begin{equation}
D(m,l)= A_{l}\  e^{\frac{-(m-l)^2}{2 w^2}} \mbox{ where} \ A_{l} =(1/ \sum\limits_{k} e^{\frac{-(k-l)^2}{2 w^2}}).
\end{equation}
In this model, we {\color{black} quantify} the "weakness" of the measurement by the quantity $w$. 
In the extreme limit of a "strong" measurement, when $w\rightarrow 0$, $D$ becomes sharp, $D(m,l) \rightarrow \delta(m-l)$, and hence, $E_{m}=\Pi_{m}$ (Fig.\ref{POVM-Figure}). In the limit of a "very weak" measurement when $w\rightarrow \infty$, $D(m,l)$ becomes a uniform distribution and hence $E_{m} \propto \mathbb{1}$, and so, the state $\rho_{0}$ is not affected by the state update rule. {\color{black} In other words, the weakest measurement {\color{black} causes} the least disturbance to the system.} \\\\
Consider {\color{black}an} $\epsilon$-polarizing {\color{black}quantum map}, $\Lambda[\rho]= (1-\lambda)\ \rho + \lambda \ \rho_{0}$, {\color{black}for} the evolution {\color{black}process} which tends to return the state to the thermal equilibrium polarization with $Tr[\textbf{J}_{z}.\rho_{0}]= \epsilon$. As an example, consider the initial state   
$\rho_{0}= \sum\limits_{k} q_{0}(k)\ \frac{\Pi_{k}}{Tr[\Pi_{k}]}$ in which $q_{0}(k)$ is a density function {\color{black} with mean value $\epsilon= \sum\limits_{k} k \ q_{0}(k)$}. {\color{black}For instance}, in the case of {\color{black}a mixed state}, $\rho_{0}=\frac{\mathbb{1}}{2^{N}}$, $q_{0}(k)=Tr[\Pi_{k}]/2^{N}$ is a binomial distribution {\color{black} with zero mean. Given $\rho_{0}$}, the first {\color{black} weak} measurement results in $m_{1}$ with {\color{black} a} probability 
\begin{eqnarray}
P(m_{1};t_{1})&=& Tr[E_{m_{1}}.\rho_{0}]\\\nonumber
&=& \sum\limits_{k} D(m_{1},k)\  q_{0}(k)\nonumber.
\end{eqnarray}
 It is known that given the {\color{black} distribution}, $P(m)$, the updated density matrix is not uniquely determined in case of a weak measurement, \cite{Nielson}. {\color{black} This is b}ecause, the set of $\{ \mathcal{M}_{m}\}$ that satisfies $\mathcal{M}_{m}^{\dagger}\mathcal{M}_{m} =E_{m}$ is not unique.  Nevertheless, one of the possible ways of updating the density matrix is $\mathcal{M}_{m}= \sqrt{E_{m}}$ {\color{black}, which gives} 
\begin{eqnarray}\nonumber
\rho^{|m_{1}}&=&\frac{ \sqrt{E_{m_{1}}}.\rho_{0}.\sqrt{E_{m_{1}}}
}{P(m_{1};t_{1})}\\\nonumber
&=& \sum\limits_{k}   q_{1}(k|m_{1}) \frac{ \Pi_{k}}{Tr[\Pi_{k}]}\nonumber.
\end{eqnarray}
Here we define $q_{1}(k|m_{1}):=\frac{ D(m_{1},k)\  q_{0}(k)}{P(m_{1};t_{1})}$ to be the updated {\color{black} density function}. As desired,  the updated density matrix collapses not only to $\frac{\Pi_{m}}{Tr[\Pi_{m}]}$ but also to other neighbor{\color{black}ing} subspaces, $k\neq m_{1}$, and its range depends on {\color {black}the} measurement "weakness" $w$. This semi-localized state around $m_{1}$, will then evolve under the $\epsilon$-polarizing {\color {black} map}, $\Lambda$. Similar to the PVM case, by performing the second measurement, we obtain a conditional distribution 
\begin{eqnarray}\nonumber
P(m_{2};t_{2}|m_{1})&=& Tr[E_{m_{2}}.\Lambda[\rho^{|m_{1}}]]\\ \nonumber
&=& (1- \lambda)\ \sum\limits_{k} D(m_{2},k)  \ q_{1}\ (k|m_{1}) \\ \nonumber
 &+&\lambda \ \sum\limits_{k} D(m_{2},k) q_{0}(k)  \nonumber.
\end{eqnarray}
The fact that the overlap between $D(m_{2},l)$ and $D(m_{1}, l)$ appears in the first term of {\color{black}the} last equation confirms that, as long as $\lambda \neq 1$ and $w\neq \infty$, there are correlations carrying on from one measurement to another.\\\\
 We calculated the joint probability distribution between any two data points and obtained
\begin{eqnarray}\nonumber
P(m_{i};t_{i}, m_{j};t_{j})&=& (1-\lambda) ^{i-j}  \sum\limits_{l} D(m_{i},l) \ D(m_{j},l)\ q_{0}(l) \\ \nonumber
&+& \eta_{i-j} \ P(m_{i};t_{i}) \  \ P(m_{j};t_{j}).
\end{eqnarray} 
{\color{black}Despite the fact that a strong and a weak measurement result in different statistical distributions, ( i.e. $Tr[\Pi_{m} \rho] \neq Tr[E_{m} \rho]$) there are common features in both limits{\color{black}, most importantly}, the statistics of instances are correlated with previous data. These correlations are a result of the quantum evolution map between measurements.}\\ \\ 
{\color{black}Thus far}, we have not included the suggested Gaussian model for $D(m,l)$. If we do so, the covariance function {\color{black} becomes}
\begin{equation}
R(k)= \frac{N}{4}\ (1-\lambda)^{k}  + (\eta_{k}-1) \ E(\textbf{M};t_{1})^{2}.
\end{equation}
One can test this relation for a totally mixed input state and reproduce the exact result in Eq.(\ref{PVMCorr}). This indicates that spin fluctuations have {\color{black} a} similar behavior in both the strong and the weak measurement limit.\\\\


\section{Conclusion}
An open quantum system model of {\color{black} the} spin noise signal in NMR was described. We have shown that the inherent spin fluctuations {\color{black} can be described by} the nature of quantum measurements,  {\color{black} the} state update rule and quantum evolution. We analyzed our model for arbitrary initial state{\color{black}s} including the identity, {\color{black}as well as} any arbitrary quantum evolution CPTP map acting on non-interacting spins{\color{black}, with the} depolarizing {\color{black} map} as an example of a collective quantum evolution. We calculated the joint probability distribution and the covariance function for different examples in both {\color{black} the} limits of strong and weak measurement.\\\\
The proposed spin noise model predicts the statistical fluctuation of {\color{black}a} spin ensemble by considering a collective measurement and a collective quantum evolution {\color{black}while retaining} the average properties such as thermal polarization. Previous computational models of spin noise have introduced a fluctuating field over the ensemble to create dephasing and account for noise correlations (\cite{QuantumOrigin}).
{\color{black} Here,} the model does not require such a field, the fluctuations are a function of the update rule that propagates over knowledge of the system. {\color{black}This analysis is intended to illustrate that with a description of the spin, the cavity and the bath interactions one may straightforwardly calculate the properties of spin noise, including its correlation function. Such descriptions are useful in analyzing experimental instances of spin noise, {\color{black}in particular, with the development of spin based quantum information processors that have long lived spin states and small number of spins.}\\\\
The authors thank O. Moussa and M. Mirkamali for helpful discussions. We acknowledge support from Industry Canada, CERC, NSERC, CIFAR and Province of Ontario.\\\\

{\color{black}
\appendix

\section{}
\label{}
In this appendix, we give a concrete example of a spin noise model. Consider $100$ spin half particles each oriented randomly in the Bloch sphere, $\rho_{0}= (\frac{\mathbb{1}}{2})^{\otimes N}$. At $t=t_{1}$ we measure the magnetization along the $z$ axis, so, the measured value $m$, is the z component of the ensemble's magnetization, i.e, $m= \sum\limits_{i=1}^{100} s_{i}$ with $s_{i}= \pm \frac{1}{2}$. Thus, $m \in \{ -50,-49, ..., 0, ..., 49, 50\}$ is a random number sampled from the probability distribution $P(\textbf{M}; t_{1})$. In this example,
\begin{eqnarray}
P_{str}(\textbf{M};t_{1})&=& Tr[\Pi_{\textbf{M}}.\rho_{0}]= \text{Bin}(N, \frac{N}{2}+ \textbf{M}, \frac{1}{2}),\\ \nonumber
P_{wk}(\textbf{M};t_{1})&=& Tr[E_{\textbf{M}}.\rho_{0}]= \sum\limits_{k=-\frac{N}{2}}^{\frac{N}{2}} D(\textbf{M}, k)\\ \nonumber
  && \hskip 2cm \times \text{Bin}(N, \frac{N}{2}+ k, \frac{1}{2}).\\ \nonumber
\end{eqnarray} 
where the subscript $st$ (or $wk$) refers to the strong (or the weak) measurement. If one repeats this first measurement with the same initial state, $\rho_{0}$, many times, a statistical distribution of $\textbf{M}(t_{1})$ will be obtained. We implemented this numerically and the results are shown in Fig.\ref{fig5}.
\begin{figure}[h]
\includegraphics[trim={6cm 0cm 0cm 0cm},width=0.6 \textwidth]{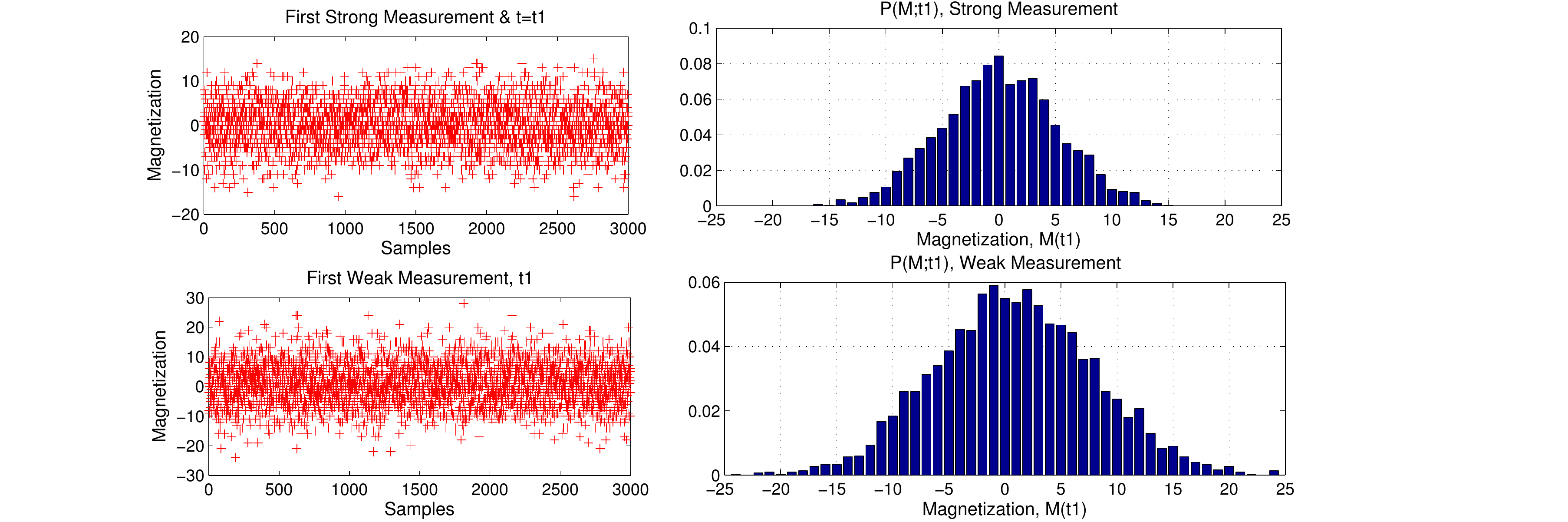}
\caption{\label{fig5}{\color{black} 3000 identical measurements are performed on 3000 identically prepared spin states with $\rho_{0} = (\frac{\mathbb{1}}{2})^{\otimes N}$. Each data point is a random number $m_{1}$ sampled from the statistical distribution $P(\textbf{M};t_{1})$. A numerical estimation of $P(\textbf{M};t_{1})$ is computed and plotted on the right side for both strong and weak measurements. $N=100$ and $w=5$.}}
\end{figure}\\
Once the data $m_{1}$ is recorded, the ensemble's density matrix is updated to $\rho^{|m_{1}}=\frac{\Pi_{m1}}{2^{N} P_{st}(m_{1};t_{1})}$ or $\rho^{|m_{1}}=\frac{E_{m1}}{2^{N} P_{wk}(m_{1};t_{1})}$ depending on whether the measurement was strong or a weak. Now, we let the spin system evolves for certain time $\delta t$ under the following quantum evolution map,

\begin{eqnarray}
\Lambda[\rho]&=& \Phi^{\otimes N}[\rho]\hskip 0.5 cm \text{where} \\ \nonumber
\Phi[\varrho]&=& (1-\lambda) U.\mathcal{\varrho}.U^{\dagger} + \lambda \frac{\mathbb{1}}{2}.
\end{eqnarray}
The CPTP map $\Phi$ acts on individual spins and for this example, we chose it to be a depolarizing map (relaxation) followed by a unitary rotation around the $x$ axis, i.e, $U= e^{- i \omega S_{x} \delta t}$. $\Phi$ is parametrized by $\lambda = 1- e^{-\delta t/T}$ and $ \theta =\omega \ \delta  t$. Following the discussion in section.\ref{subA}, the probability of spin flip is $\alpha = \beta = (1-\lambda)(\sin \theta)^{2} + \frac{\lambda}{2}$. Following the spin noise model, once the system is measured and evolves under $\Lambda$, the second measurement is performed at $t=t_{2}$. The second measured outcome $m_{2} \in \{ -50,-49, ..., 0, ..., 49, 50\}$ is again a random number sampled from the conditional probability distribution $P(\textbf{M}; t_{2}| m_{1})$. For this particular example, we computed the conditional probability distribution in the case of the strong measurement,

\begin{eqnarray}
P_{st}(\textbf{M}; t_{2}| m_{1})&=& Tr[\Pi_{\textbf{M}}. \Lambda[\rho^{|m_{1}}] ]\\ \nonumber
&=& \sum\limits_{i=0}^{\frac{N}{2}+ m_{1}} \text{Bin}(\frac{N}{2}+ m_{1}, i, \alpha)\\ \nonumber
        &\times &\text{Bin}(\frac{N}{2}- m_{1}, \textbf{M}-m_{1}+i, \alpha)
\end{eqnarray} 
 as well as for the weak measurement,
\begin{eqnarray}
P_{wk}(\textbf{M}; t_{2}| m_{1})&=& Tr[E_{\textbf{M}}. \Lambda[\rho^{|m_{1}}] ]\\ \nonumber
&=& \sum\limits_{k, k'=-\frac{N}{2}}^{\frac{N}{2}} D(\textbf{M},k) D(m_{1},k')\text{Bin}(N, \frac{N}{2}+ k', 1/2)\\ \nonumber
        &\times & \frac{P_{st}(k;t_{2}|k')}{P_{wk}(m_{1};t_{1})}
\end{eqnarray}\\
For the numerical simulation, we prepared identical initial states $\rho_{0}$, performed the first measurement on them, then post selected on that data with magnetization value, $\textbf{M}=m_{1}$. Given, these selected conditional states, $\Lambda[\rho^{|m_{1}}]$, we made a second measurement and recorded the data and its statistics as shown in Fig.\ref{fig6}. 
\begin{figure}[h]
\center
\includegraphics[trim={5cm 0cm 0cm 0cm}, width=0.65 \textwidth]{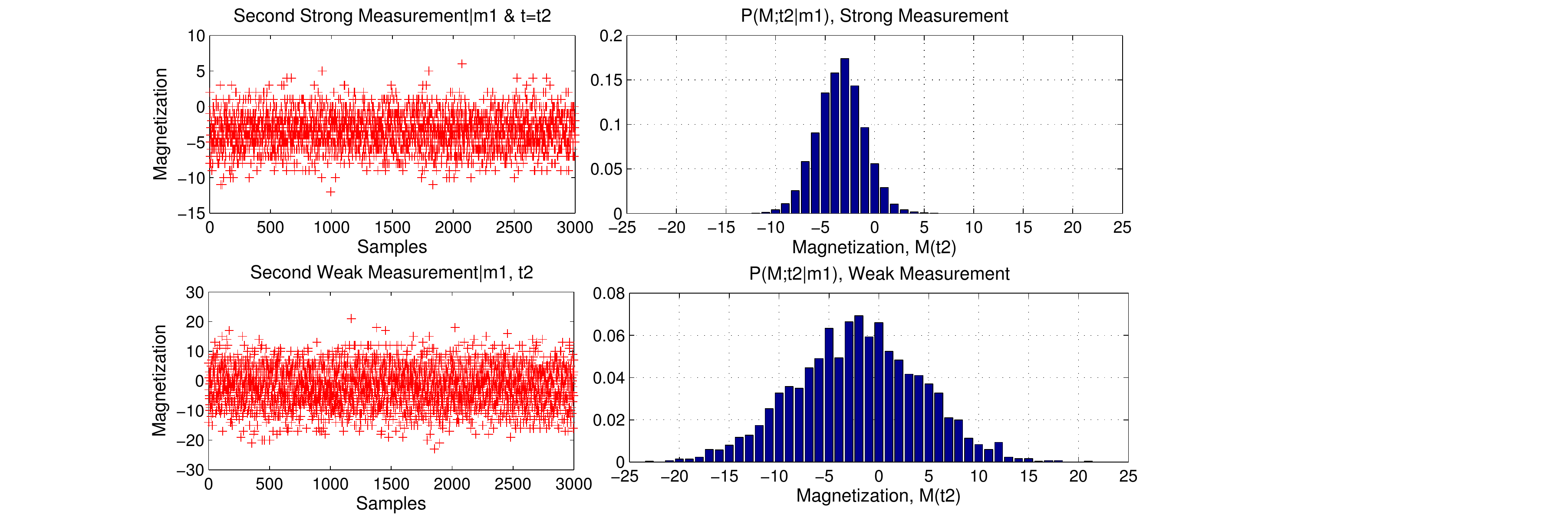}
\caption{\label{fig6} {\color{black} The raw data (left) and its corresponding statistical distribution (right) of the second measurement is presented for both cases of the strong and the weak measurements. Here, $\lambda= 0.1$ and a small unitary rotation, $\theta=\pi/32$, are considered. }}
\end{figure}\\
 To see the correlation between the two subsequent data points, we plot the joint probability distribution $P(\textbf{M}_{2}, \textbf{M}_{1}) = P(\textbf{M}_{2}|\textbf{M}_{1}) P(\textbf{M}_{1})$ where $\textbf{M}_{i}= \textbf{M}(t_{i})$. As shown in Fig.\ref{fig7}, for the case of $\lambda=0$, $\theta=0$ (no evolution) and strong measurement, there is a maximum correlation between the two data points.
\begin{figure}[h]
\center
\includegraphics[trim= {1cm 0.5cm 2cm 0cm}, width=0.6 \textwidth]{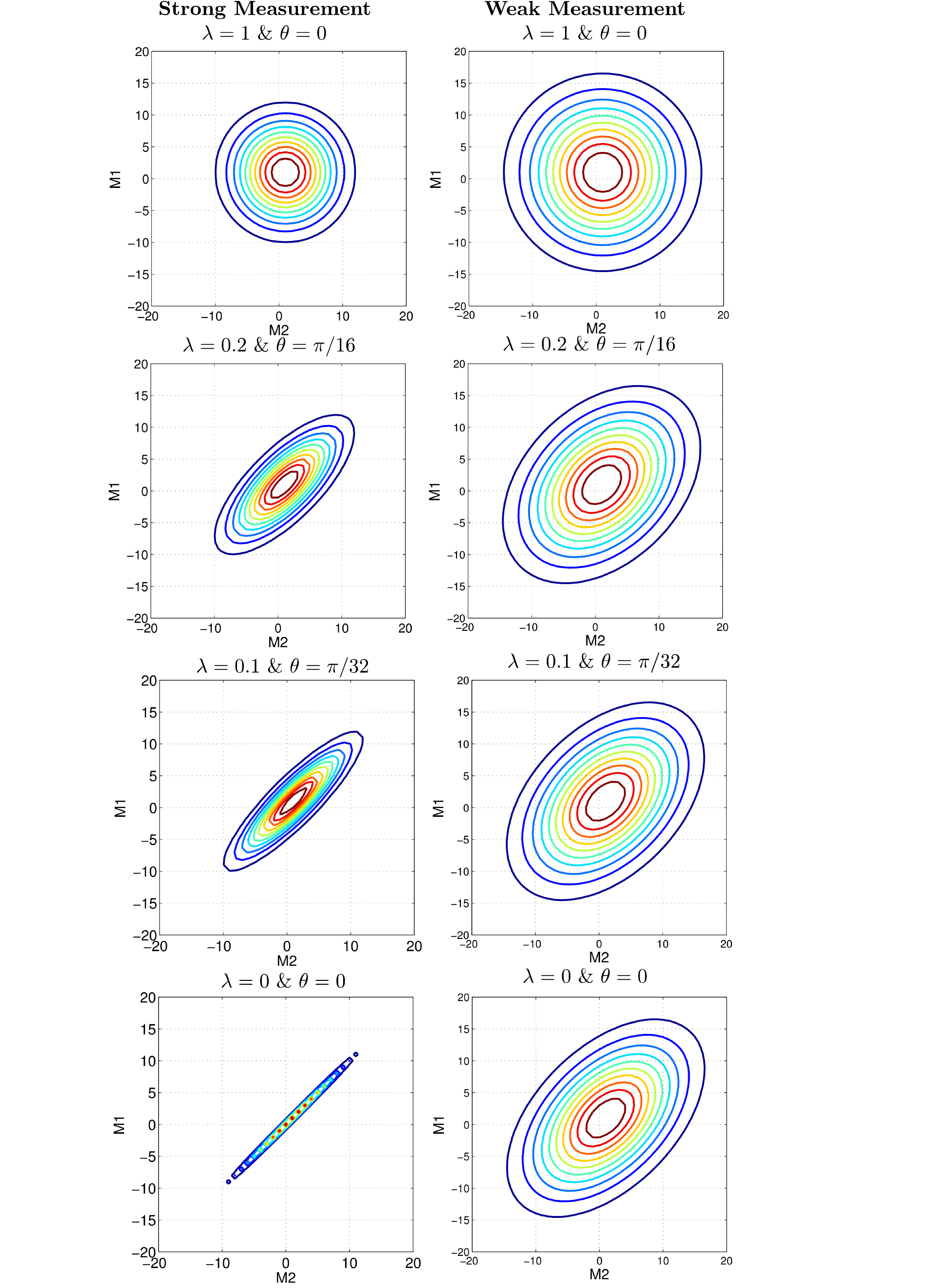}
\caption{\label{fig7} {\color{black} Contour plots of the joint probability distribution, $P(\textbf{M}_{2},\textbf{M}_{1})$ for different values of $\lambda$ and $\theta$. The top figure $\lambda=1$ and $\theta=0$, shows no correlation between subsequent data points and the bottom figure $\lambda=0$ and $\theta=0$ shows a maximum correlation between them.}}
\end{figure}
 But, for the case of weak measurement, there is less correlation even in the absence of any evolution. This shows that the data may not be reproducible in the case of a weak measurement. As the spin flip probability, $\alpha$, becomes larger (longer evolution), the two data points become less and less correlated as seen in Fig.\ref{fig7}.\\\\
Measurement of subsequent data points proceeds in exactly the same style and the correlation function neighboring points does not change. Correlations are observed in all of these cases with the weak measurement data having less correlation than the strong.}\\\\


\end{document}